\documentstyle[aps,preprint,epsfig]{revtex}

\begin{document}

\title{Total-energy-based prediction of a quasicrystal structure}

\author{M. Mihalkovi\v{c}$^{1,2}$, I. Al-Lehyani$^3$, E. Cockayne$^4$,
C.L. Henley$^5$, N. Moghadam$^6$, J.A. Moriarty$^7$,
Y. Wang$^8$, M. Widom$^3$}

\address{
$^1$ Institute fur Physik, Technische Universit\"{a}t Chemnitz, 
D-09107 Germany \\
$^2$ Institute of Physics, Slovak Academy of Sciences, Bratislava, Slovakia \\ 
$^3$ Department of Physics, Carnegie Mellon University, Pittsburgh PA  15213 \\
$^4$ Ceramics Division, NIST, Gaithersburg MD 20899-8520 \\
$^5$ Department of Physics, Cornell University, Ithaca NY  14853 \\
$^6$ Oak Ridge National Laboratory, Oak Ridge TN 37831-6114 \\
$^7$ Lawrence Livermore National Laboratory, Livermore CA 94551 \\
$^8$ Pittsburgh Supercomputer Center, Pittsburgh PA  15213 \\
}
\date{\today}

\maketitle

\begin{abstract}
Quasicrystals are metal alloys whose noncrystallographic symmetry
and lack of structural periodicity challenge methods of experimental
structure determination.  Here we employ quantum-based total-energy
calculations to predict the structure of a decagonal quasicrystal from
first principles considerations. We employ Monte Carlo simulations,
taking as input the knowledge that a decagonal phase occurs in
Al-Ni-Co near a given composition, and using a few features of the
experimental Patterson function.  The resulting structure obeys a
nearly deterministic decoration of tiles on a hierarchy of length
scales related by powers of $\tau$, the golden mean.
\end{abstract}

\newpage

Al-Ni-Co forms thermodynamically stable and highly perfect decagonal
quasicrystalline samples over a range of
compositions~\cite{PhaseDiagram}. Of special interest is the
composition Al$_{0.70}$Ni$_{0.21}$Co$_{0.09}$ for which the structure
is periodic along the z axis with a period of $c=4.08$~\AA, and
quasiperiodic perpendicular to this axis with a characteristic length
(termed a ``quasilattice constant'') of $a_0=2.45$~\AA~\cite{basicNi}.
This ``basic Ni'' composition is well suited for theoretical modeling
because it should be the simplest structure, lacking the quasiperiodic
modulation and $c$-axis doubling observed at other compositions.
Numerous attempts to determine the structure of this compound start
from experimental data~\cite{Patterson,AlNiCoModels,Z-contrast} but do
not {\it predict} a structure on the basis of total energy.

The transition metals Ni and Co (generically denoted ``TM'') play
similar chemical roles in Al-transition metal quasicrystals, and they
are not distinguished by ordinary X-ray or electron diffraction. Our
model predicts distinct sites for Ni and Co near the ``basic''
composition.

Our total energy calculations employ quantum-based pair-potentials
derived from the generalized pseudopotential theory (GPT)~\cite{mGPT}.
GPT expands the total energy in a series of volume-, pair- and
many-body-potentials. The volume term exerts no force and may be
neglected at fixed volume and composition. The many-body terms are
small except among clusters of neighboring transition-metal atoms, and
we incorporate their influence with modified short-ranged TM-TM pair
interactions constrained by full {\em ab initio} calculations.

We identify four salient properties of the computed oscillatory
potentials~\cite{mGPT}: (1) V$_{\rm AlAl}(r)$ has a broad shoulder
starting around $r = 2.9$~\AA~ and is repulsive at shorter distances;
(2) V$_{\rm AlCo}(r)$ and V$_{\rm AlNi}(r)$ exhibit deep first minima
near $r=2.5$~\AA~ and second minima near $r=4.5$~\AA; (3) The V$_{\rm
AlCo}$ well is significantly deeper than the V$_{\rm AlNi}$ well; (4)
The modified $V_{\rm TMTM}(r)$ have shallow minima near r=2.6~\AA.

\begin{figure}
\vspace{-0.6 in}
\centerline{\epsfig{file=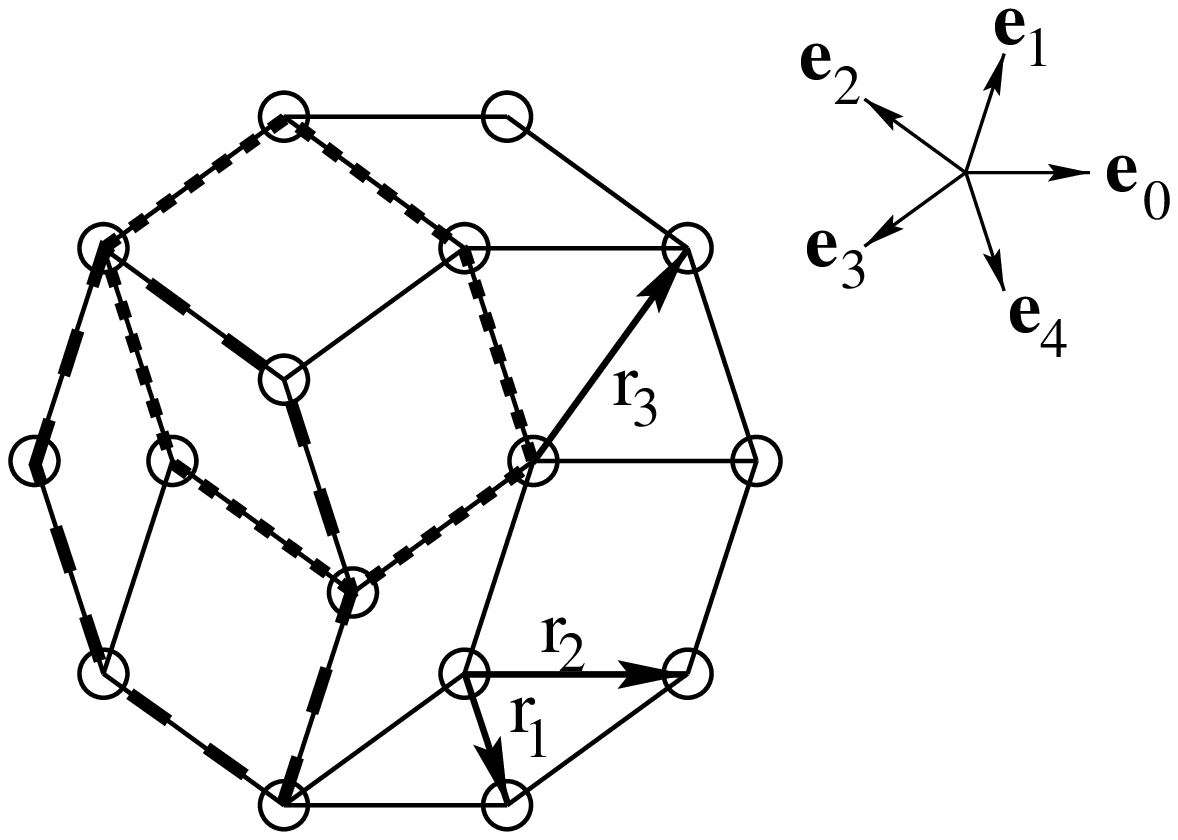,width=5in}}
\vspace{-1 in}
\caption{Random rhombus tiling decorated with ideal sites. The
long- and short-dashed lines outline, respectively, thin and fat
hexagons.  Unit vectors $\{{\bf e}_i\}$ lie parallel to tile
edges. Vectors ${\bf r}_i$ are defined in the table.}
\label{layer}
\end{figure}

The following features of the $d$(AlNiCo) structures are evident in
the experimentally determined Patterson function~\cite{Patterson}
which contains a peak at every interatomic vector $\bf r$: (A) All
atoms lie on or nearly on layers separated by $c/2=2.04$~\AA; (B) The
vector from an atom to a nearest neighbor (with a tolerance of $\sim
0.1$~\AA) belongs to a small, discrete basis set of ``linkage''
vectors; (C) The in-plane components of linkage vectors are $\pm a_0
{\bf e}_i$ (see Fig.~\ref{layer}) or simple sums of such vectors.

We construct trial quasicrystal structures that achieve low total
energy while satisfying the above experimental constraints. To enforce
constraints (B) and (C) we limit atomic positions to a collection
of discrete sites (Fig.~\ref{layer}), located at vertices of a
two-dimensional tiling of rhombi with edge $a_0$ and acute angles
$36^\circ$ or $72^\circ$. To enforce constraint (A) we stack two
independent tilings above each other.  As the tiles can be placed in
many ways, and atoms distributed randomly among the sites, these {\it
minimal constraints} permit a great variety of structures, including
all reasonable quasicrystal structures. After we discover favorable
low energy motifs consistent with the minimal constraints, we
remove unnecessary degrees of freedom, effectively defining
{\it highly constrained} models.

A Metropolis Monte Carlo annealing yields low energy structures.  Two
kinds of Monte Carlo steps are employed: (i) swaps between nearby
atoms of different species in either layer, including hops of one atom
to an empty ideal site nearby; (ii) ``flips'' which reshuffle the
three rhombi in a fat or thin hexagon, in one layer. (see
Fig.~\ref{layer}) Due to the hexagon flips, our ensemble is an
``equilibrium random tiling'' allowing phason disorder~\cite{Phason},
but the system is free to find a quasiperiodic state if that is
favored by the potentials. Our simulations are performed with periodic
boundary conditions using cell sizes chosen to best approximate
the quasiperiodic structure.

\begin{figure}[!ht]
\centerline{
\epsfig{file=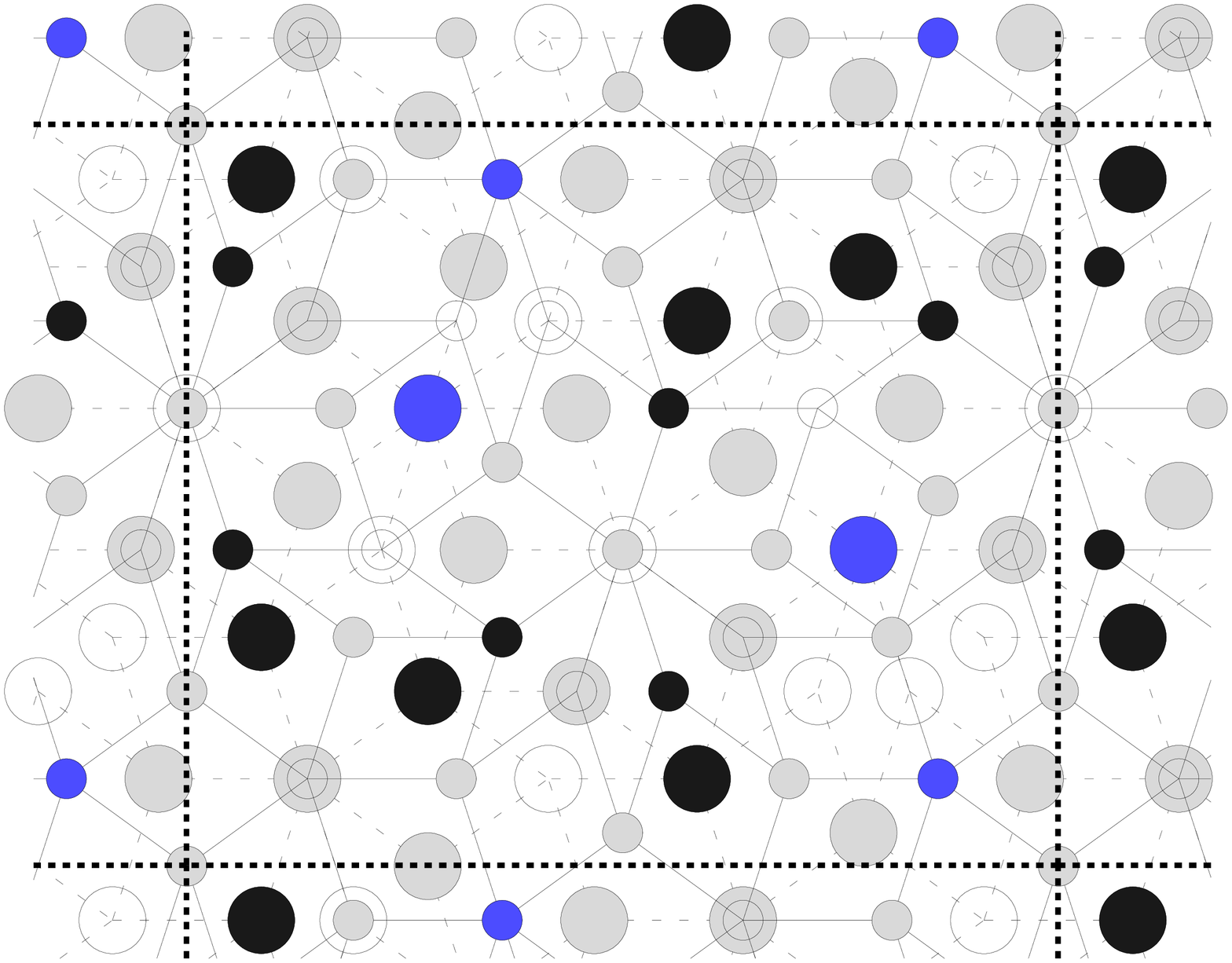,width=4in,angle=90}}
\vspace{-15pt}
\centerline{
\epsfig{file=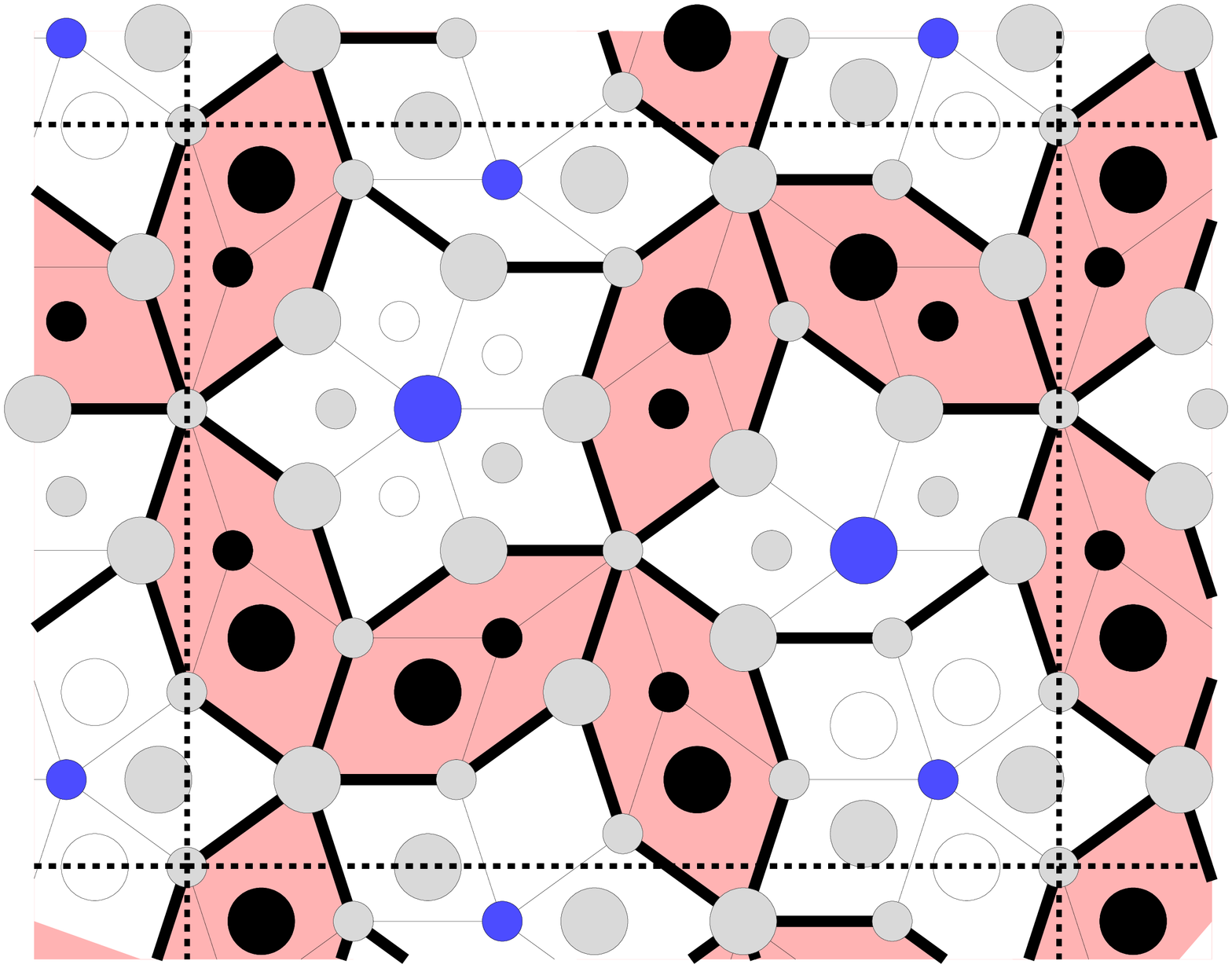,width=4in,angle=-90}}
\caption{Minimum energy configurations. Small/large circles indicate atoms in
upper/lower layer. Gray=Al, Blue=Co, Black=Ni, White=vacant. (a) Top
figure results from minimally constrained simulation (solid/dashed
lines denote upper/lower tilings). (b) Bottom figure results from highly
constrained simulation. Dark solid lines outline $a_0$-scale HBS
tiling.  Pink-shaded hexagons connect vertices of $\tau^2 a_0$-scale HBS
tiling.}
\label{small}
\end{figure}

In our initial simulation with minimal constraints, we employ a cell
of size $12.22 \times 14.37 \times 4.08$~\AA$^3$ and composition
Al$_{34}$Ni$_{12}$Co$_{4}$.  This cell contains 72 ideal sites, 36 in
each layer with the $c$-axis periodicity enforced. Slow cooling
identifies a unique minimum energy configuration illustrated in
Fig.~\ref{small}a.

The optimal configuration can be described simply in terms of a new
{\it highly constrained} model (see Fig.~\ref{small}b) that obeys the
following rules: (i) the entire plane is tiled by three compound tiles
called ``hexagon'', ``boat'', and ``star'', outlined by heavy edges
and built respectively of three, four, and five rhombi (this is called
the ``HBS'' tiling~\cite{CW}).  (ii) The optimal decoration of the HBS
tiles is virtually unique.  Minimally constrained simulations with
larger cells support these rules.

We can understand this decoration in terms of the salient
features of the potential we enumerated at the start.  In view of features
(1) and (2), the minimum-energy structure must maximize the number of
Al-Co and Al-Ni bonds.  In view of feature (3), every Co ought to have
purely Al neighbors, which is geometrically feasible just up to
$\sim$10~\% Co, which is the ``basic'' composition.  Thus, all TM-TM
neighbors must be Ni-Ni. Every Ni has mostly Al neighbors but cannot
escape having two or three Ni neighbors, since $\sim$30~\% of all atoms
are transition metals.

Let us check which ideal-site separations are favorable for which atom
pairs.  Within the same layer, the tile edge length $r_2$=2.45~\AA~ is
unfavorable for Al-Al or TM-TM bonds, but highly favorable for Al-TM
bonds. However, because of the high density of Al atoms, we find a
small number of Al-Al bonds do take this length. The short diagonal of
a fat rhombus is $r_3=2.88$~\AA, which is an acceptable Al-Al
distance. Hence the $72^\circ$ Al-TM-Al isosceles triangle (half a fat
rhombus) is highly favored within a layer.

The interlayer spacing $c/2$=2.04~\AA~ is too short for any pair.
Sites in adjacent layers, spaced by $\tau^{-1} a_0$ in-plane (e.g.,
the short diagonal of the thin rhombus), are separated by $r_1'=2.54$~\AA~
which is favorable for Al-TM or TM-TM bonds.  Finally, sites in
adjacent layers separated by $a_0$ in-plane have a total separation of
$r_2'=3.19$~\AA~ which is an acceptable Al-Al distance.

Given this understanding of chemical bond lengths we can easily
justify the decoration of the HBS tiles. Each tile is bounded by Al
atoms of alternating heights at separation $r_2'$. Interior sites of
the hexagon tile are too close to the Al border for Al atoms. Since it
must hold two TM atoms, it holds a pair of Ni atoms at distances
$r_1'$ and $r_2$ from the border Al and mutual separation $r_1'$.
Four of the border Al atoms form a rectangle with edges $r_2'$ and $r_3$
lying in a plane that is nearly the perpendicular bisector of the
Ni-Ni bond. This fragment of the hexagon tile is thus a slightly
distorted region of B2 (CsCl) structure~\cite{Steurer}.

The interior vertex of the boat and star tiles are at the ideal TM
distance $r_2$ from border Al atoms. Since this is a point of high Al
coordination, it is occupied by Co.  The boat and star tiles have room
for two additional interior Al atoms at separation $r_3$.  In an
isolated star tile this interior Al pair can lie in any of five
symmetry-related configurations. The structure surrounding the star
generally breaks this degeneracy by means of long-range interactions.

In the decoration just described, an Al atom on an HBS tiling vertex
is often at the center of a small cluster which was an important motif
of earlier models~\cite{Burkov,CW}. This cluster appears in
Fig.~\ref{small}b wherever hexagons join at their tips.  This cluster
consists of a pentagon of mixed Al and Ni atoms in the same layer as
the vertex Al, and additional pentagons in the adjacent layers above
and below. These adjacent pentagons contain only Al atoms and are
rotated by 36${\circ}$ with respect to the middle pentagon.
This cluster exhibits interlayer Al-Ni separations of 2.54~\AA~ and
4.46~\AA, precisely at the first and second minima of V${\rm AlNi}$.

Since the HBS tile corners are all multiples of $72^\circ$, edges
emanating from the HBS corner atoms in one layer can only point in the
five directions $+{\bf e}_i$ while those within the other layer point
in the directions $-{\bf e}_i$.  Statistically, the layers are
equivalent but related by a screw axis.  Allowing for the reflection
planes normal to the layers, and in the absence of further
symmetry-breaking, the HBS decoration implies a space group
$10_5$/mmc, consistent with experiment.

For the next level of modeling, we take the highly constrained HBS
tiles as fundamental objects.  Tile-tile interactions are defined
implicitly as the sum of the pair potentials between atoms decorating
the tiles.  The allowed ``flips'' (Monte Carlo moves) of the HBS
tiling are called ``bow tie flips'' as the tile edges before and after
the flip outline a bow tie shape~\cite{CW}.  The bow tie flips are
generated by fat hexagon flips of the underlying rhombus tiling.
Additionally the Al pair inside the star can rotate among its five
allowed orientations.  The reduced degrees of freedom make the highly
constrained HBS tiling much faster to simulate at low temperatures
than the minimally constrained rhombus tiling.

The ensemble of random HBS tilings contains a variable tile frequency
ratio H:B:S because HS pairs interchange with BB pairs by bow tie
flips. Highly constrained simulations forbid this flip because it
alters the chemical composition, given our ideal tile decoration.
There is a particular ``golden'' ratio H:B:S=$\sqrt{5}\tau:\sqrt{5}:1$
that is obtained, for example, by removing double-arrow edges from a
Penrose tiling. Decorating such a tiling deterministically yields an
ideal composition Al$_{0.700}$Ni$_{0.207}$Co$_{0.093}$ and atomic
volume 14.16~\AA$^3$. Both composition and atomic volume are
consistent with experiment.

\begin{figure}[!ht]
\centerline{
\epsfig{file=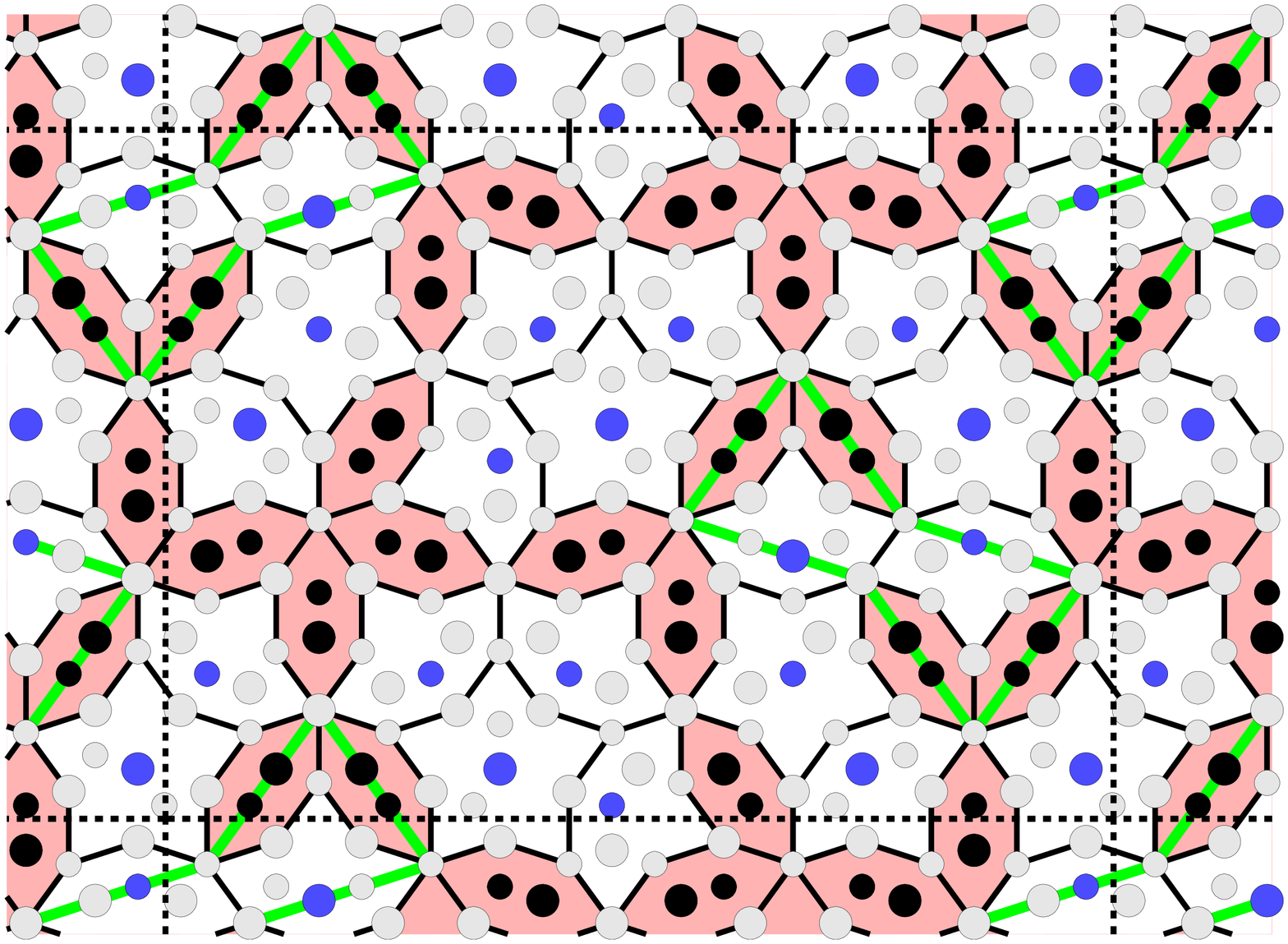,width=4in,angle=-90}}
\vspace{-15pt}
\centerline{
\epsfig{file=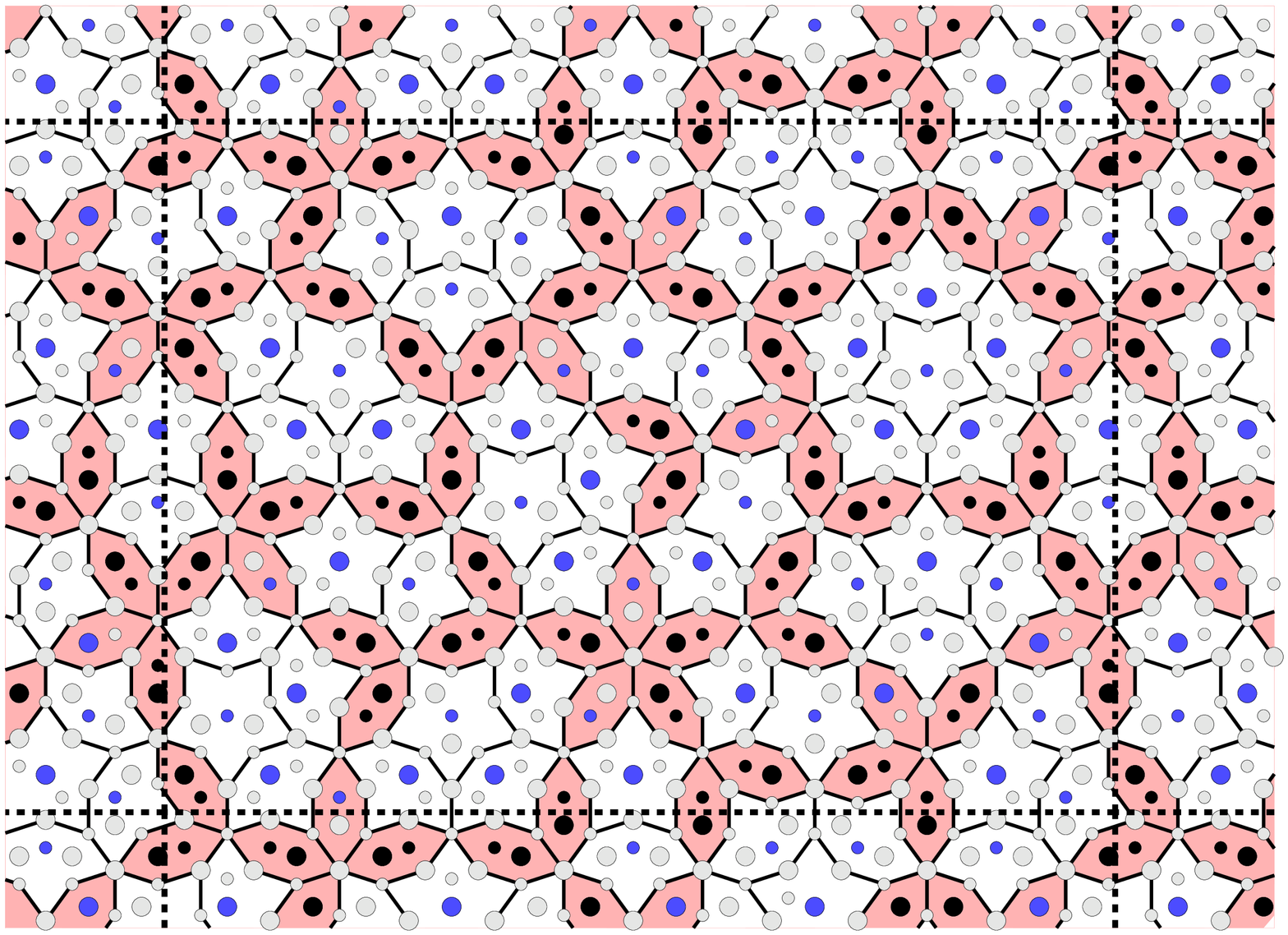,width=4in,angle=-90}}
\caption{Lowest energy configurations obtained.
(a) Top shows highly constrained simulation. Green lines outline bow
tie tiles. (b) Bottom variable occupancy simulation.}
\label{large}
\end{figure}

Large-scale simulations (see Fig.~\ref{large}) reveal a ``supertile''
ordering in which hexagons connect tip-to-tip.  Each hexagon
tip becomes a vertex of a ``supertiling'', with longer edges of
length $\tau^2 a_0$ along the midline of every hexagon.  Since
orientations of adjoining hexagons differ by multiples of $72^\circ$
degrees, the same is true for their midlines, hence the ``supertile''
edges differ by $72^\circ$ angles and form mainly HBS tiles (as well
as a new ``defect'' tile).

The supertile atomic structure is mechanically stable.  Under relaxation of
the structure shown in figure ~\ref{small} (which consists of two large
scale hexagons), the average Co and Ni displacement is just
0.10~\AA. The average Al displacement is 0.17~\AA~ except for the two
Al atoms located slightly off-center in the large-scale hexagons which
displace 1.13~\AA~ to the symmetric points at the hexagon centers.

The ``defect'' tile breaks the connectivity of the small-hexagon
chain, introducing a new tile shape which we call a ``bow tie''.  To
understand the role of the bow tie and its low symmetry decoration,
consider the energetics of the large scale HBS tilings.  Because the
large-scale HBS tile decoration is essentially deterministic, we may
replace the actual interatomic interactions with effective
interactions between tiles. We find a single parameter dominates the
energetics: an energy cost is associated with 72$^{\circ}$ junctions
between tile edges decorated with Ni atom pairs, because the resulting
high density of Ni atoms reduces the number of favorable Al-Ni bonds.
Defect tiles enter only when they reduce the number of 72$^{\circ}$
junctions.  They accomplish this reduction by interchanging a NiNi
pair with a nearby AlCo pair.

An alternate means of reducing the frequency of 72$^{\circ}$ angles
between tile edges decorated with Ni atom pairs is to alter the
chemical composition. To accommodate the new composition we must relax
certain constraints in our simulation. We keep the small scale
HBS tiles with Al fixed on their boundaries, but allow arbitrary
chemical occupancy of the interior sites.  Replacing 20~\% of NiNi
pairs with AlCo pairs eliminates all defects from the minimum energy
configuration and leads to chemical composition
Al$_{0.720}$Ni$_{0.166}$Co$_{0.114}$, still within the limits of the
basic Ni composition. The low energy configurations consist entirely
of HBS tiles decorated as found previously, but now many tile edges that
participate in two 72$^{\circ}$ junctions get decorated with an AlCo
pair rather than a NiNi pair (see Fig.~\ref{large}b).

We compare our model with experimental Z-contrast electron microscope
imagery~\cite{Z-contrast}. This experimental method images atomic
columns proportionally to the mean square atomic number of the column,
so the images translate quite directly into TM positions. A key
feature of the experimental data is the occurrence of decagonal rings
with a 20~\AA~ diameter. These rings exhibit up to 10 TM doublets
around the perimeter, an interior ring of 10 TM singlets, and a
central triangular core~\cite{Z-contrast}. This characteristic
structure is in excellent agreement with our model, where two hexagons
and a boat frequently coalesce into a decagonal cluster (see center of
Fig.~\ref{large}b). The triangular core of this cluster, which breaks
decagonal symmetry, is recognized as the sail of the boat tile. Full
{\it ab-initio} calculations~\cite{Yan} recently verified energetic
favorability of this particular core structure.

Our study began with interatomic potentials plus a minimum of
experimental information. We derived structural models starting with a
minimally constrained lattice gas on a fluctuating rhombus tiling.
Systematically removing unnecessary degrees of freedom yielded a
nearly deterministic decoration of HBS tiles at a length scale
$\tau^2$ times larger than the initial rhombus tiling, a model
consistent with Z-contrast electron microscopy.  This procedure can be
repeated to identify yet larger characteristic atomic clusters
providing a novel example of multiscale modeling which might be
applicable to other quasicrystals.

\thanks{This research was supported by the National Science Foundation 
and the Department of Energy.}

\newpage

\begin{table}[!ht]
\begin{tabular}{|l|l|l|}
\hline
\hline
Label           & Example                              & Comments \\
\hline
$r_1$=1.51~\AA  & $a_0({\bf e}_0+{\bf e}_3$)           & Forbidden   \\
$r_2$=2.45~\AA  & $a_0 {\bf e}_0$                      & Al-TM, (Al-Al) \\
$r_1'$=2.54~\AA & $a_0({\bf e}_0+{\bf e}_3)+{{c}\over{2}}\hat{z}$ 
                                                       & Al-TM, TM-TM  \\
$r_3$=2.88~\AA  & $a_0({\bf e}_1-{\bf e}_0)$           & Al-Al \\
$r_2'$=3.19~\AA & $a_0 {\bf e}_0+{{c}\over{2}}\hat{z}$ & Al-Al \\
\hline
\hline
\end{tabular}
\caption{Characteristic distances (illustrated in Fig.~\ref{layer}) 
and important bond types. Primed vectors connect adjacent layers.}
\label{tab1}
\end{table}

\newpage

\newpage

\newpage

\begin{thebibliography}{99}

\bibitem{PhaseDiagram} 
Ritsch, S., {\it et al.}, {\it Phil. Mag. Lett.} {\bf 78},67-76
(1998).

\bibitem{basicNi} 
Ritsch, S., {\it et al.}, {\it Phil. Mag. Lett.}, {\bf 74}, 99-106
(1996); Tsai, A.P., Fujiwara, A., Inoue, A. and Masumoto, T., {\it
Phil. Mag. Lett.}, {\bf 74} 233-40 (1996).

\bibitem{Patterson} 
Steurer, W., Haibach, T., Zhang, B., Kek, S. and L\"uck, R., {\it Acta
Cryst. B} {\bf 49}, 661-75 (1993).

\bibitem{AlNiCoModels}
Yamamoto, A. and Weber, S., {\it Phys. Rev. Lett.}, 4430-3 {\bf 78}
(1997); Steinhardt, P.J., {\it et al.}, {\it Nature} {\bf 396}, 55-7
(1998); Kraj\v c\'i, M., Hafner, J. and Mihalkovi\v c, M., {\it Phys.
Rev. B} {\bf 62}, 243-55 (2000).

\bibitem{Z-contrast} 
Abe, E., {\it et al.}, {\it Phys. Rev. Lett.} {\bf 84} 4609-12 (2000);
Yan, Y., Pennycook, S.J.  and Tsai A.P., {\it Phys. Rev.  Lett.} {\bf
81}, 5145-8 (1998).

\bibitem{mGPT} 
Moriarty, J.A. and Widom, M., {\it Phys. Rev. B} {\bf 56}, 7905-17
(1997); Al-Lehyani, I., {\it et al.}, cond-mat/0010205, submitted to
{\it Phys. Rev. B} (2000).

\bibitem{Phason} 
Henley, C.L., ``Random tiling models'' in {\it Quasicrystals: The
state of the art}, ed. D.P. DiVincenzo and P.J.  Steinhardt, (World
Scientific, Singapore, 1991) p. 429

\bibitem{Burkov} 
Burkov, S.E., {\it Phys. Rev. Lett.} {\bf 67}, 614-17 (1993).

\bibitem{CW} 
Cockayne, E. and Widom, M., {\it Phys. Rev. Lett.} {\bf 81}, 598-601
(1998).

\bibitem{Steurer}
W. Steurer, {\it Mat. Sci. Eng.} A {\bf 294-296}, 268-271 (2000).

\bibitem{Yan}
Y. Yan and S.J. Pennycook, {\it Phys. Rev. Lett.} {\bf 86}, 1542-1545 (2001).

\end{thebibliography}
\enddocument